\documentclass[12pt]{article}
\pdfoutput=1

\usepackage[toc,page]{appendix}
\usepackage{slashed}
\usepackage[sumlimits,intlimits,namelimits]{amsmath} 
\usepackage[english]{babel}
\usepackage{graphicx}
\usepackage{amssymb}
\newcommand{\be}{\begin{equation}}
\newcommand{\ee}{\end{equation}}
\newcommand{\ba}{\begin{eqnarray}}
\newcommand{\ea}{\end{eqnarray}}

\newcommand{\al}{\alpha}
\newcommand{\alp}{\frac{\alpha}{\pi}}
\newcommand{\MeV}{{\rm\,MeV}}
\newcommand{\GeV}{{\rm\,GeV}}
\newcommand{\ice}[1]{\relax}

\usepackage{color}
\usepackage{amsmath}
\usepackage{booktabs} 
\usepackage{subfigure}

\begin{document}

  
\begin{flushright}
SI-HEP-2019-14\\
SFB-257-P3H-19-037
\end{flushright}
\begin{center}
{\Large\bf Hadronic contributions to the muon anomalous moment}
\end{center}
\begin{center}
{\sc Stefan Groote$^1$}, {\sc Thomas Mannel$^2$}, and
{\sc Alexei A. Pivovarov$^2$}\\[0.1cm]
{\sf$^1$ F\"u\"usika Instituut, Tartu \"Ulikool,
W.~Ostwaldi 1, 50411 Tartu, Estonia}\\[0.3cm]
{\sf$^2$ Theoretische Elementarteilchenphysik, Naturwiss.-techn. Fakult\"at, \\
Universit\"at Siegen, 57068 Siegen, Germany}
\end{center}
\begin{abstract}\noindent
We discuss the hadronic contributions to the muon anomalous magnetic moment. 
They are dominated by light quark contributions which are constrained by the 
mechanism of chiral symmetry breaking. Using the leading order result based on 
$e^+ e^-$ scattering data, we show that the next-to-leading order contributions 
in the fine structure constant
$\alpha$ can be reliably calculated. Extending this idea to the hadronic 
four-point function we give a prediction for the light-by-light 
contribution.  
\end{abstract}


\section{Introduction\label{sec:Intro}}
The prediction of a $g$ factor $g_e = 2$ for the magnetic moment of
the
electron
marked a great success of the relativistic wave equation introduced by Dirac
in 1928~\cite{Dirac:1928hu}. With Schwinger's pioneering analysis of the
electron magnetic moment within perturbation theory
in 1948~\cite{Schwinger:1948iu}, Quantum
Electrodynamics (QED) as
the first quantum field theory was established. Presently
the lepton anomalous magnetic moments continue to be important observables for
precision tests of the Standard Model (SM)~\cite{Kinoshita:1990nb}. 
Current data indicate a tension with the theoretical prediction for the 
anomalous magnetic moment of the muon which is considered to be 
the most promising observable for a stringent test of the SM and
for searches for beyond SM physics (as a review, see e.g.\
Ref.~\cite{Jegerlehner:2017gek}). The recent experimental value for the
anomalous magnetic moment of the muon is~\cite{Bennett:2006fi}
\begin{equation}
g_\mu/2=1+a_\mu=1.001\,165\,920\,8(6).
\end{equation}
The Particle Data Group (PDG) gives an updated value for the muon anomaly
in the form~\cite{PDG}
\begin{equation}
a_\mu^{\rm exp} = 116\,592\,091(54)(33)\times 10^{-11}.
\end{equation}
The current muon experiment at Fermilab plans to reduce the experimental
uncertainty by the factor of four~\cite{exp2019},
$\sigma_{\rm future}\approx (1.0\div 1.5)\times 10^{-10}
=(10\div 15)\times 10^{-11}$. This precision clearly is a challenge for the 
theoretical side to increase the precision of the prediction. 

The theoretical results for the muon anomalous magnetic moment in the SM are 
traditionally represented as a sum of three parts,
\begin{equation}
\label{eq:a-SM}
  a_\mu^{\rm SM}=a_\mu^{\rm QED}+a_\mu^{\rm EW}+a_\mu^{\rm had}
\end{equation}
with $a_\mu^{\rm QED}$, $a_\mu^{\rm EW}$ being the leptonic and electroweak parts,
respectively, and $a_\mu^{\rm had}$ is the contribution involving the electromagnetic 
currents of quarks. We note that in
eq.~(\ref{eq:a-SM}) $a_\mu^{\rm EW}$ also contains 
the quark loops related to the interaction with the heavy week bosons. Since their
contribution is small, we do not worry about a precise description of this hadronic
contribution to $a_\mu^{\rm EW}$. 

The leptonic part is computed in perturbation theory and reads~\cite{PDG}
\begin{equation}
a_\mu^{\rm QED}=116\,584\,718.95(0.08)\times 10^{-11}.
\end{equation}
The computation extends up to five-loop level, using both analytical and
numerical techniques~\cite{tenthQED} (as a review see e.g.\
Ref.~\cite{qedcont}). At present, the numerical results are steadily being
checked/refined with powerful analytical methods for Feynman integral
evaluation. In view of the experimental uncertainty, the QED part of the
theory prediction for the muon anomaly can be considered to be exact, giving
a negligible uncertainty.

The electroweak part is known to two loops and reads~\cite{PDG} 
\begin{equation}
a_\mu^{\rm EW} = 153.6(1.0)\times 10^{-11}\, .
\end{equation}
The absolute value of $a_\mu^{\rm EW}$ is small and the uncertainty of this
contribution is
negligible, at least for comparison with the present experiments.

The hadronic part $a_\mu^{\rm had}$ in the SM is related to quark contributions
to the electromagnetic currents. 
To leading order (LO) in the fine structure constant $\alpha$ it is given by
the two-point function of hadronic electromagnetic currents through the vacuum
polarization of the photon. In order to match the experimental accuracy of the
muon anomaly one has to include next-to-leading order (NLO) contributions in
$\alpha$. At this order the four-point function of hadronic electromagnetic
currents starts to contribute.

As we shall discuss below, an accurate calculation of the hadronic 
contributions due to light quarks is very difficult as they are represented 
by (almost) massless light quarks and are
infrared (IR) singular in perturbation 
theory. This is the main obstacle for obtaining high precision SM predictions.
Instead, the theoretical estimate for the hadronic two-point function utilizes 
scattering data. 
The LO hadronic contribution extracted from $e^+e^-$ data is given
by~\cite{Davier:2017zfy} 
\begin{equation}\label{eq:pdgLO}
a_\mu({\rm LO};{\rm had};e^+e^-)=6931(33)(7)\times 10^{-11}.
\end{equation}
Other estimates are based on data from hadronic
$\tau$ lepton decays~\cite{Jegerlehner:2017gek} and yield
\begin{equation}
a_\mu({\rm LO};{\rm had};\tau)=(6894.6\pm 32.5)\times 10^{-11}.
\end{equation}
In our estimates we stick to the PDG value in Eq.~(\ref{eq:pdgLO}) for
definiteness, called $a_\mu({\rm LO};{\rm had})$.

The hadronic contribution is rather large and needs to be computed with a
precision of one or two per mille. This is a challenge for the theory, since 
there are no appropriate tools for an analytical theoretical
computation. However, presently the lattice is emerging as a promising tool for this
task. 

In NLO there are further hadronic contributions. They are extensively
discussed in the literature and estimated in various approaches. The current
total SM prediction reads~\cite{PDG}
\begin{equation}
a_\mu^{\rm SM}=116\,591\,823(1)(34)(26)\times 10^{-11}.
\end{equation}
The difference
\begin{equation}
\Delta a_\mu = a_\mu^{\rm exp}-a_\mu^{\rm SM}=268(63)(43)\times 10^{-11}
\end{equation}
could be due to physics beyond the SM, but it is not statistically
significant yet; however, it is considered to be rather serious for the prospect of
discovering new physics. 

The main theoretical uncertainties originate from hadronic contributions. 
These are encoded in the two-point function and the four-point function of 
the hadronic electromagnetic currents. The four-point function is
involved in the topology called light-by-light (LBL)~\cite{PDG}.
In the present paper we re-consider these hadronic contributions. The method 
to deal with the genuinely nonperturbative light-quark contributions makes 
use of the mechanism of chiral symmetry breaking which is assumed to capture 
the main physics to describe the hadronic matrix elements involving light 
quarks. This is described in the next section. 
We use this idea to update the results
of Refs.~\cite{Pivovarov:2001mw,Groote:2001vu} 
and calculate the light-by-light
contribution in this approach. Some useful formulas are 
given in the Appendix.

\section{Description of the
  method\label{sec:technique}}
One of the key features of massless QCD is spontaneous breaking of chiral 
symmetry, which determines the low-energy behavior of the lightest states 
of QCD. Some recent discussion and references can be found in Ref.~\cite{Mannel:2019grm}. 
This spontaneous symmetry breaking (SSB) results at the level 
of correlation functions in the generation of mass terms 
for light fermion propagators in the complex (exact) QCD vacuum, i.e.\
the light quark is not just the Lagrangian quark anymore but rather a dressed
collective excitation.

In perturbation theory the chirality of massless quarks is conserved, which 
can be read off from the perturbative light quark propagator in Fock space, 
\begin{equation}\label{eq:light-quark-Fockq}
S(q)=\frac{1}{\slashed{q}}\sim \frac{1}{\slashed{x}x^2}
\end{equation}
However, the (nonperturbative) interaction with soft gluons changes this behavior 
at large distances $x$, i.e. for small values of $q$  (cf.\ e.g.\ Ref.~\cite{Korchemsky:1985ts}). Using the operator
product expansion (OPE) as proposed by Wilson~\cite{Wilson:1969zs}, one finds the expansion (cf.\ e.g.\ Ref.~\cite{Chetyrkin:1988yr})
\begin{equation} \label{prop1}
S(q) = \frac{\slashed{q}}{q^2} + c_{qq} \frac{\langle \bar{q} q \rangle}{q^4} 
+  c_{GG} \frac{\slashed{q}}{q^2}  \frac{\langle G^2  \rangle}{q^4} + \cdots
\end{equation} 
where $\langle\bar q q\rangle$ is the quark condensate, $\langle G^2  \rangle$ is 
the gluon condensate and $c_{qq}$, $c_{GG}$ are Wilson coefficients which can be 
computed in perturbation theory. 

The physical meaning of the condensate term is that an effective mass term for the 
light quarks emerges through chiral SSB, which can be taken into account by writing 
\begin{equation}\label{eq:light-quark-Ext}
S(q,A)=(\slashed{q}-M(q))^{-1} \, \, \mbox{with} \,\, 
M(q) = c_{qq}\frac{\langle \bar{q} q \rangle}{q^2}   
\end{equation}
for the quark two-point point function in the presence of gluons. 
Here the dynamical mass $M(q)$ describes the effects of
chiral symmetry breaking at the level of Green functions (cf.\ e.g.\
Ref.~\cite{Anselm:1975ka}), and the relation given in
eq.~(\ref{eq:light-quark-Ext}) 
is the asymptotic form of $M(q)$ for large $q^2$.  
The exact asymptotic behavior at short distances can
be obtained in OPE through the quark condensate as order
parameter~\cite{Politzer:1976tv}, or from the Dyson--Schwinger integral
equation in the spirit of self-consistency or gap equations familiar from
superconductivity~\cite{Krasnikov:1982gp}. 

In fact, this approach becomes more transparent by starting from the functional-integral 
expression for the two-point function for the hadronic electromagnetic current
\begin{equation}\label{eq:light-quark-two-point-latt}
\Pi_2(x)=\int [DA]\, {\rm Tr} [S(x,A)\gamma^\mu S(-x,A)\gamma_\mu]
\end{equation}
where $\int [DA]$ represents the functional integration over the gluon field  
with a proper weight. Furthermore, $S(x,A)$ is the light-quark propagator in the 
presence of gluon field in the coordinate space. As a side remark we
note,
that the expression (\ref{eq:light-quark-two-point-latt})
is the starting point for a lattice calculation
of the two-point function for the 
hadronic electromagnetic current. 

The expression~(\ref{eq:light-quark-two-point-latt}) is genuinely
nonperturbative, so there 
is no way to perform an actual analytical calculation.
In our approach we assume that the major 
effect of the integration over the gluon fields is the breaking of chiral symmetry which amounts
to replace the light-quark propagator by (\ref{eq:light-quark-Ext}) such that  
\begin{equation}\label{eq:light-quark-two-point-mass}
\int [DA] \,  {\rm Tr}[S(x,A)\gamma^\mu S(-x,A)\gamma_\mu] = 
{\rm Tr}[S(x,\hat{M}(x))\gamma^\mu S(-x,\hat{M}(x))\gamma_\mu]
\end{equation}
with the dynamical mass
$\hat{M}(x)$~\cite{Diakonov:1995ea,Dorokhov:2001wx}
which is related to 
$M(q)$ by Fourier transformation. 

The behavior of the dynamical mass as a function of $x$ (or likewise $q$) is 
only qualitatively known. However, we note that the exact integral of the two-point hadronic
function for the muon anomalous magnetic moment as in Eq.~(\ref{directcont})
given in Appendix
is equal to an integral of the function in
Eq.~(\ref{eq:light-quark-two-point-mass}) for some constant value $M(q)=m^*$, 
which is always true for reasonably smooth functions.
The entire analysis can be done in Euclidean space-time which contains no
particle singularities and where $m^*$ provides a straight
infrared cut-off of QCD.
Practically, this is a very efficient ansatz as all correlation functions are
indeed represented by Feynman diagrams for which the analytical expressions are
known.

The numerical value for the parameter $m^*$ of our ansatz,
$m^* =180\MeV$~\cite{Pivovarov:2001mw} (see
Sect.~\ref{sec:HadConNLO}), has been extracted from data of $e^+ e^-$ scattering
in LO, i.e.\ from measured hadronic two-point function. This value turns out 
to be rather close to both the pion mass and the usual constituent quark masses.  
However, this fact is purely accidental, since $m^*$ in our approach is 
simply an infrared cutoff parameter in massless QCD and specific for the considered 
observable, namely the muon anomalous magnetic moment. On the other hand, a value in this 
ballpark is to be expected, giving us some confidence that the mechanism of 
chiral SSB is indeed the main input in the  light-quark dynamics relevant for the 
anomalous magnetic moments. 

Turning now to the NLO hadronic contributions we first consider the ones which 
need the hadronic two-point function as an input. NLO in this context means to 
look at the leptonic corrections and to compute the relevant integration 
kernel $K^{(4)}$ which has a very similar shape as the LO integration kernel 
$K^{(2)}$, i.e. we have to a very good approximation~\cite{Groote:2001vu}
$$
K^{(4)} (x) \propto  \frac{\alpha}{\pi} K^{(2)} (x) 
$$
and hence we have the same convolution integral with the hadronic two-point function,
up to a constant. In the approach discussed above this means that the same value for 
$m^*$ has to be used when evaluating the NLO leptonic contributions, and the uncertainty 
of the input LO contribution then directly translates into that of the NLO 
contribution.

The extrapolation of this procedure to contributions from the hadronic
four-point function is not so obvious, since the integration weight functions
(kernels) for the anomalous magnetic moment are now of different form, which 
would lead to a different value of $m^*$ for the four-point function. However, 
if the mechanism of chiral SSB remains to be the main piece, we do not expect 
$m^*$ to be grossly different form the one extracted from the two-point function.  
The assumption, namely that $m^*$ has the same value in the calculation of the 
anomalous magnetic moment of the muon, is the main systematic uncertainty of this
approach. While this is a point difficult to resolve by analytical methods,
the explicit numerical calculations on the lattice can quantitatively test
this assumption in the future.

Although we assume that the effect of chiral symmetry breaking and the appearance 
of the quark condensate is the leading contribution, we may also consider the effect 
of the gluon condensate. To this end, we can write  
\begin{equation}
S(q,A) = R(q)(\slashed{q}-M(q))^{-1} \, \, \mbox{with} \,\, 
R(q) = 1 + c_{GG} \frac{\langle G^2  \rangle}{q^4}
\end{equation}
where the expression for $R(q)$ is again given in the asymptotic regime. Note 
that the above expression equals (\ref{prop1}) up to terms of order $1/q^7$.   
The emergence of the gluon condensate in QCD is related rather to the breaking 
of scale invariance than to the chiral symmetry, and it is not clear whether 
the gluon condensate is an order parameter of some symmetry breaking phenomenon. 

\begin{figure}\begin{center}
		\includegraphics[scale=0.3]{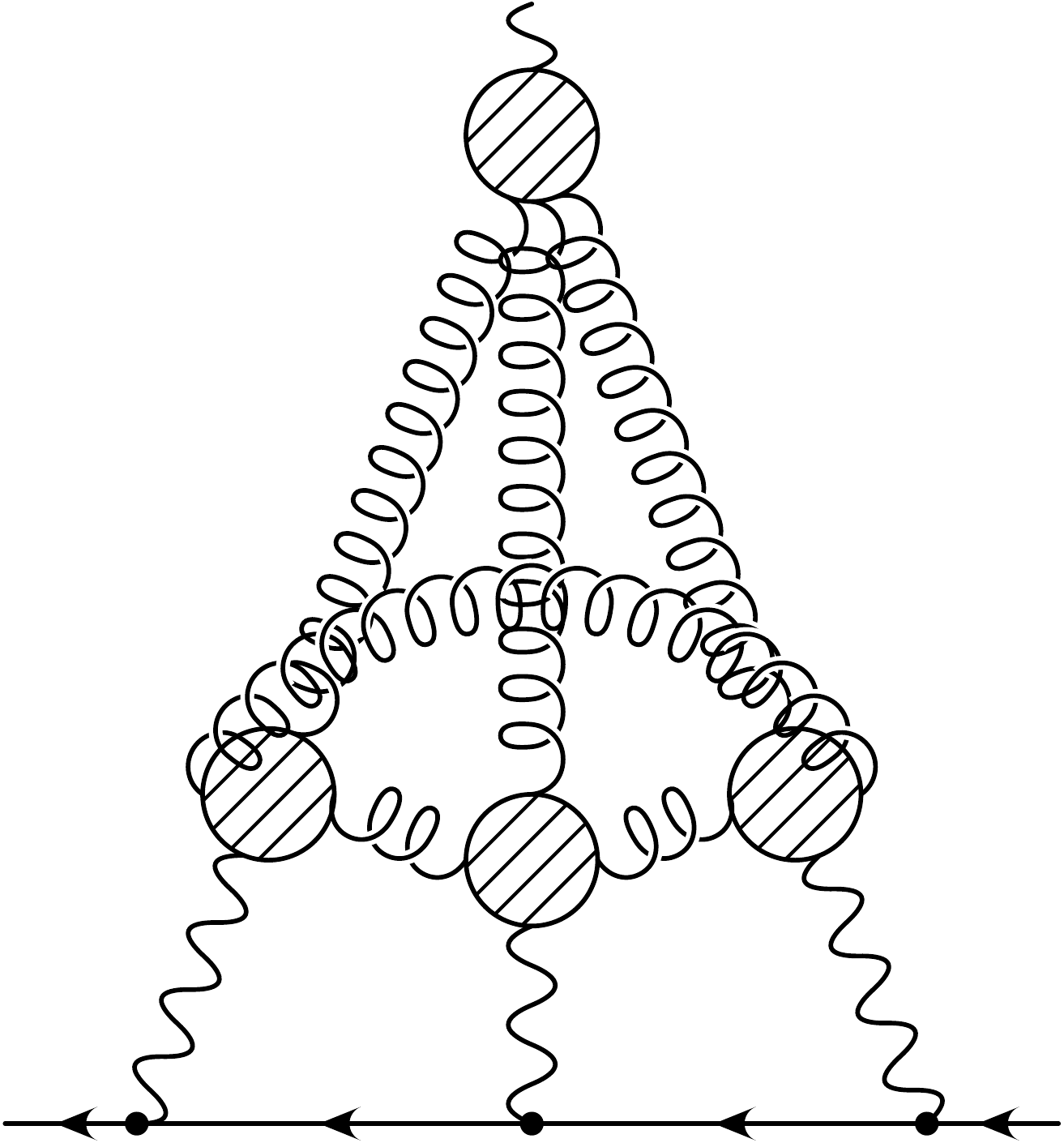}\quad
		\includegraphics[scale=0.3]{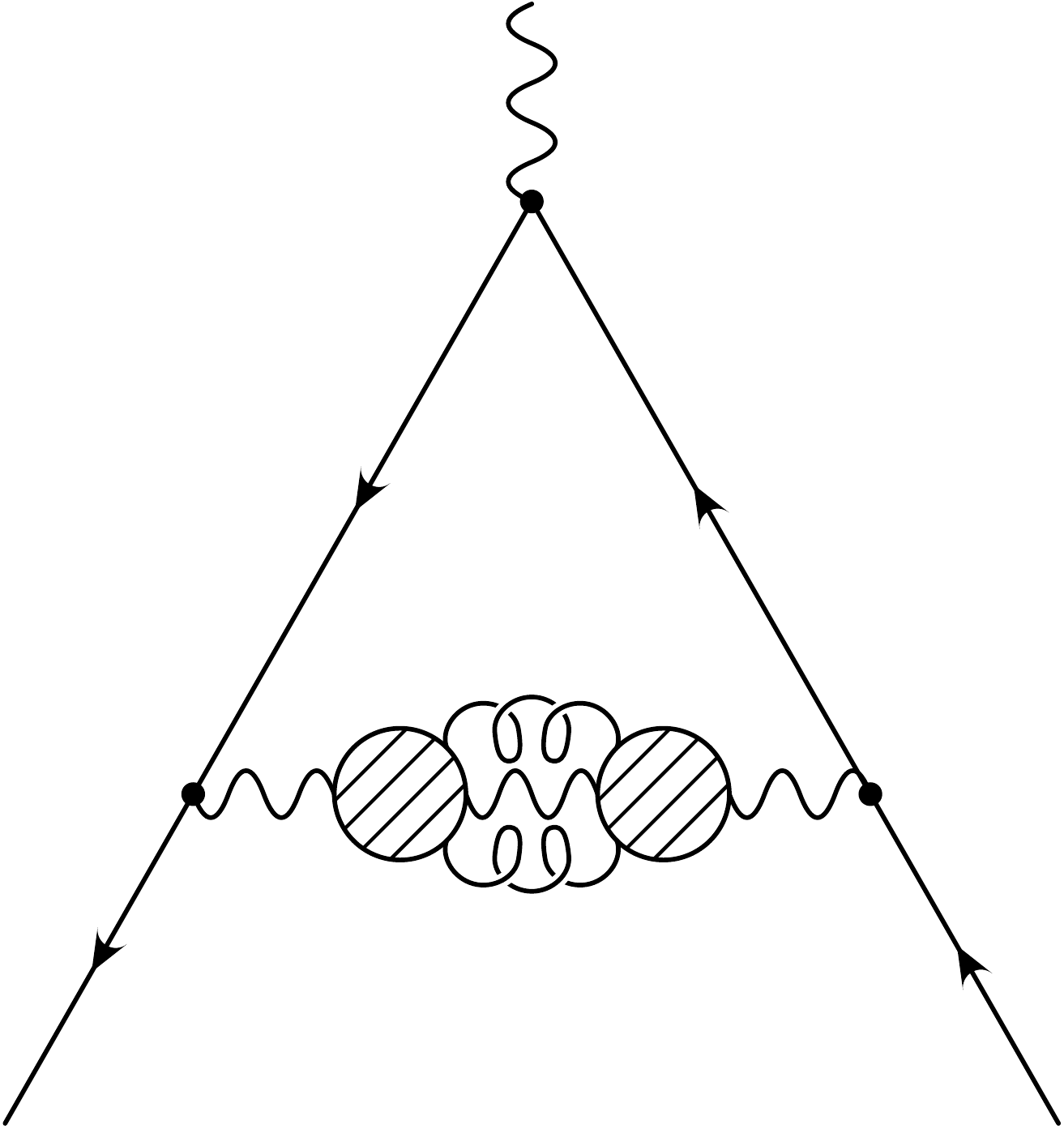}\quad
	\includegraphics[scale=0.3]{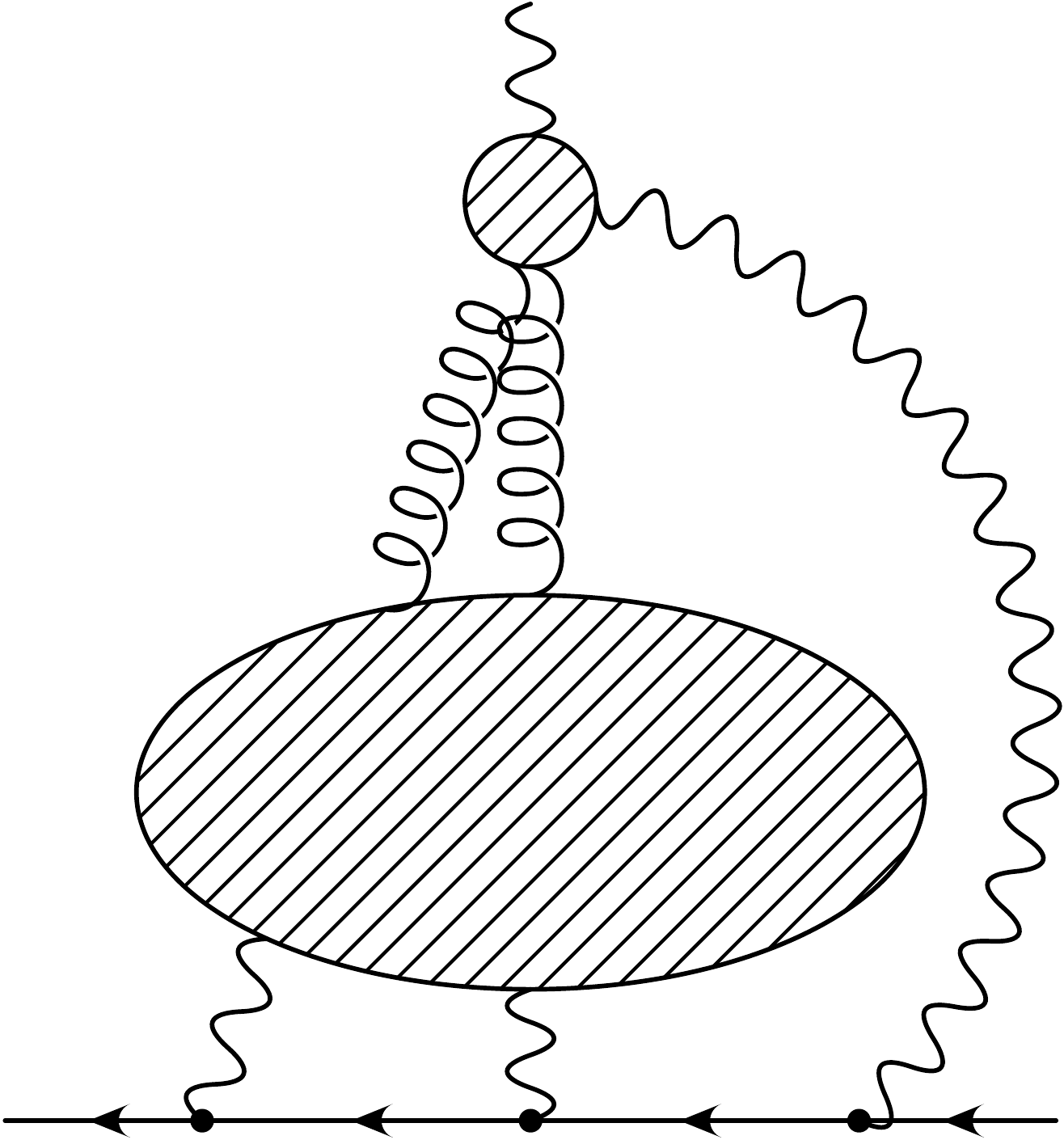}
		\caption{\label{torn}Three examples of topological
                   configurations of the four-point function}
\end{center}\end{figure}

In the same spirit as we discussed the mass $M(q)$, replacing it effectively
by a constant value $m^*$, we assume that the contributions related to $R(p)$ 
can be replaced by a constant value $r^*$. In this case all quark propagators 
will be multiplied by the constant $r^*$, which means that the hadronic 
contributions of two-point functions will be multiplied by $(r^*)^2$, while the 
hadronic contributions to the four point functions will be multiplied
by $(r^*)^4$. 
This will also modify the extraction of the value of $m^*$, since the value
extracted from the data is actually $m^*/r^*$, while the mass entering the 
four-point function would become $m^*/(r^*)^2$. 
The total contribution of the four-point function can be rather
enhanced compared to a simple picture based on perturbation theory in Fock
space. One can draw many topologically different configurations that could be
relevant for the lattice, a few examples are found in Fig.~\ref{torn}. There is
no {\it a priory\/} reason for them to be small,
however, there is also no reason 
why $r^*$ is very large
since the gluon condensate is numerically small.
Nevertheless, it is clear that within our approach
the main 
systematic uncertainty in the determination
of the hadronic contribution coming 
from the hadronic four-point function is related to the choice of the
numerical value for the effective mass $m^*$.

In the following we give
a more detailed account of the calculation and our results. 

\section{LO hadronic contributions\label{sec:HadConLO}}
To LO the hadrons contribute through a two-point function of electromagnetic
currents. In the SM this two-point function is the correlator of
electromagnetic currents of quarks (see Appendix for details). The top,
bottom, and charm quarks are heavy enough for perturbative QCD to apply.
The perturbative
corrections are given in terms of $\alpha_s(m_Q)$ and are well under control.

For a hadronic scale $m_Q$ the contribution to the muon magnetic
moment scales as $(m_\mu/m_Q)^2$. Thus,
the top quark contribution is negligible.

The bottom quark ($Q_b=-1/3$,
$m_b = 4.8\GeV$) gives 
\be\label{eq:b-LO}
a_\mu^{\rm had}({\rm LO};b)= 1.9\times 10^{-11} \, ,
\ee
where we used the pole mass~\cite{mb,Hoang:2000yr}. This contribution is
below the expected experimental uncertainty. The result~(\ref{eq:b-LO}) is
stable against the inclusion of higher order QCD corrections which are
completely negligible.

The charm quark gives a larger contribution due
to its larger electric charge and its smaller mass
  ($Q_c=2/3$, $m_c\sim 1.6\GeV \sim m_{J/\psi}/2$)
\begin{equation}
\label{eq:a-charm}
a_\mu^{\rm had}({\rm LO};c)= 69.3\times 10^{-11} \, .
\end{equation}
As stated in the introduction, the present requirement for a solid theoretical 
estimate for the anomalous 
magnetic moment of the muon is that its uncertainty should be smaller than the benchmark
uncertainty of the Fermilab experiment. This is still the case for our estimate of
the charm quark contribution. The charm quark mass is given with high
precision in~\cite{Allison:2008xk}.

Now we turn to the contribution from the light quarks. However, here a 
perturbative calculation is not possible,
since the scale for QCD corrections is 
$ \mu \sim m_q\ll\Lambda_{\rm QCD}$, and thus we use the method
described in
Sect.~\ref{sec:technique}.
We estimate the contribution of light quarks
using $e^+e^-$ data as 
\begin{eqnarray}
\label{eq:LOuds-data}
  a_\mu^{\rm had}({\rm LO}; uds-data)
  &=&
a_\mu^{\rm had}({\rm LO};{\rm all})-a_\mu^{\rm had}({\rm LO};b)
-a_\mu^{\rm had}({\rm LO};c)\nonumber \\
&=&(6931(34) - 1.9 - 69.3)\times 10^{-11}
 = 6860(34)\times 10^{-11}\, .
\end{eqnarray}
From this result the numerical value for $m^*$ is extracted as
$m^* = 180.0\pm 0.5~{\rm MeV}$.

In fact, the data-based result for the LO contribution~(\ref{eq:LOuds-data})
includes implicitly some of the NLO hadronic
corrections. These are, for instance, additional
leptonic and hadronic contributions to the vacuum polarization diagrams, or vertex
corrections
which are found in both the $e^+e^-$ data and the muon anomalous magnetic 
moment. This is a well
known problem of potential double counting, which is intensively 
discussed in the literature.
We do not consider this problem here and take the value from
Eq.~(\ref{eq:LOuds-data}) as our input for the LO part.

\section{NLO hadronic contributions
  \label{sec:HadConNLO}}
As has been discussed above, we fix the
nonperturbative effective IR mass of the light quarks using the LO value for
the hadronic contribution to the muon magnetic
moment to obtain $m^*=180\MeV$ as proposed in~\cite{Pivovarov:2001mw}. 
To be precise, computing the LO
value within our approach with $m^*=180.0\pm 0.5\MeV$ we obtain
$a_\mu^{\rm had}({\rm LO};uds)=(6852\pm 38)\times 10^{-11}$, i.e. we reproduce the 
value given in~(\ref{eq:LOuds-data}). One should not take the high precision of the
determination of the effective mass $m^*$
too seriously, since the main uncertainty of
our approach is systematic, see the discussion of the method in
Sect.~\ref{sec:technique}.  However, the structure for hadronic correlators is
completely fixed in our approach and, with the value of the infrared
mass
$m^*$ known from LO, we have an explicit model for the NLO calculations.
This model can be easily applied, since all necessary formulas are well 
known in the literature; the main source for the analytical results results 
used here is~\cite{barbieri}.

As has been pointed out before, we have at NLO  contributions involving the hadronic two-point 
as well as the hadronic four-point function, which will be considered in the subsequent subsections.

\subsection{Two-point function and photon--muon corrections}
The first contribution is given by the vertex of the type $K^{(4)}$ in
Ref.~\cite{barbieri} (cf.\ also Ref.~\cite{samuel}).
Using the expression for the kernel $K^{(4)}$ from Ref.~\cite{barbieri} and our approximation 
for the quark propagators, we find for the NLO contribution
\[
a_\mu^{\rm had}({\rm ver};{\rm NLO};uds) = - 188\times 10^{-11}\, .
\]
We note that this contribution can be also obtained by an expansion in inverse powers of the quark mass, which in our case will be the effective mass $m_q=m^*$. The relevant 
ratio is in this case $m_\mu^2 / (4 m^{*\, 2})$ since the threshold is actually at 
$2 m^*$. This expansion yields (without the QCD color factors)
\begin{equation}
a_\mu^{\rm had}({\rm ver};{\rm NLO},q)
=-\frac{8}{3}\left(\frac{m_\mu}{2 m_q}\right)^2\left(
-\frac{2689}{5400}+\frac{\pi^2}{15}
+\frac{23}{90}\ln\frac{m_q}{m_\mu}\right)\left(\alp\right)^3
\end{equation}
Inserting the QCD color factors and $m_q = m^*$ we get (cf.~\cite{kinohad})
\begin{equation}
\label{vermassexp}
a_\mu^{\rm had}({\rm ver};{\rm NLO};uds)=-171\times 10^{-11}\, .
\end{equation}
From this results
we conclude that the expansion in inverse powers of $m^*$ yields already a 
pretty precise prediction. 

The charm quark and bottom quark contributions can be calculated
perturbatively; for the charm quark we get the result 
\begin{equation}
a_\mu^{\rm had}({\rm ver};{\rm NLO};c)=-4\times 10^{-11}\, .
\end{equation}
while the bottom quark contributes a tiny amount 
\begin{equation}
a_\mu^{\rm had}({\rm ver};{\rm NLO};b) = - 0.1\times 10^{-11} \, . 
\end{equation}
Both contributions are smaller than the expected 
uncertainty of a new experimental value and
can be neglected. Obviously, the leading-order mass expansion gives sufficient
accuracy for heavy quark contributions. One can also use the expansion for the
$K^{(4)}$ kernel of Ref.~\cite{barbieri} given in Ref.~\cite{krause}, even
though the exact result is easy to handle as well.

The total vertex-type contribution reads
\begin{equation}
\label{eq:NLO-ver}
a_\mu^{\rm had}({\rm ver};{\rm NLO}) = - 192\times 10^{-11}\, .
\end{equation}

The second contribution is of the double bubble (db) vacuum polarization type
where the second 1PI block is given by leptons different from the muon, as the
muon has been already included in the vertex type contribution as
defined in~\cite{barbieri}.
The electron loop gives
\begin{equation}
\label{eq:e-uds}
a_\mu^{\rm had}({\rm db};{\rm NLO};e\& uds)=104\times 10^{-11}
\end{equation}
that should be compared to Ref.~\cite{Pivovarov:2001mw,kinohad,krause}.
The $\tau$ lepton loop is negligible,
\begin{equation}
\label{eq:tau-uds}
a_\mu^{\rm had}({\rm db};{\rm NLO};\tau\& uds)=0.05\times 10^{-11}\, .
\end{equation}
The electron loop together with a charm quark loop is marginal,
\begin{equation}
\label{eq:e-c}
a_\mu^{\rm mod}({\rm db};{\rm NLO};e\& c)=1.1\times 10^{-11}\, ,
\end{equation}
while the electron loop together with a bottom quark loop is completely negligible.

\subsection{Four-point function contributions
  \label{subsec:4point}}
\begin{figure}\begin{center}
    \includegraphics[scale=0.3]{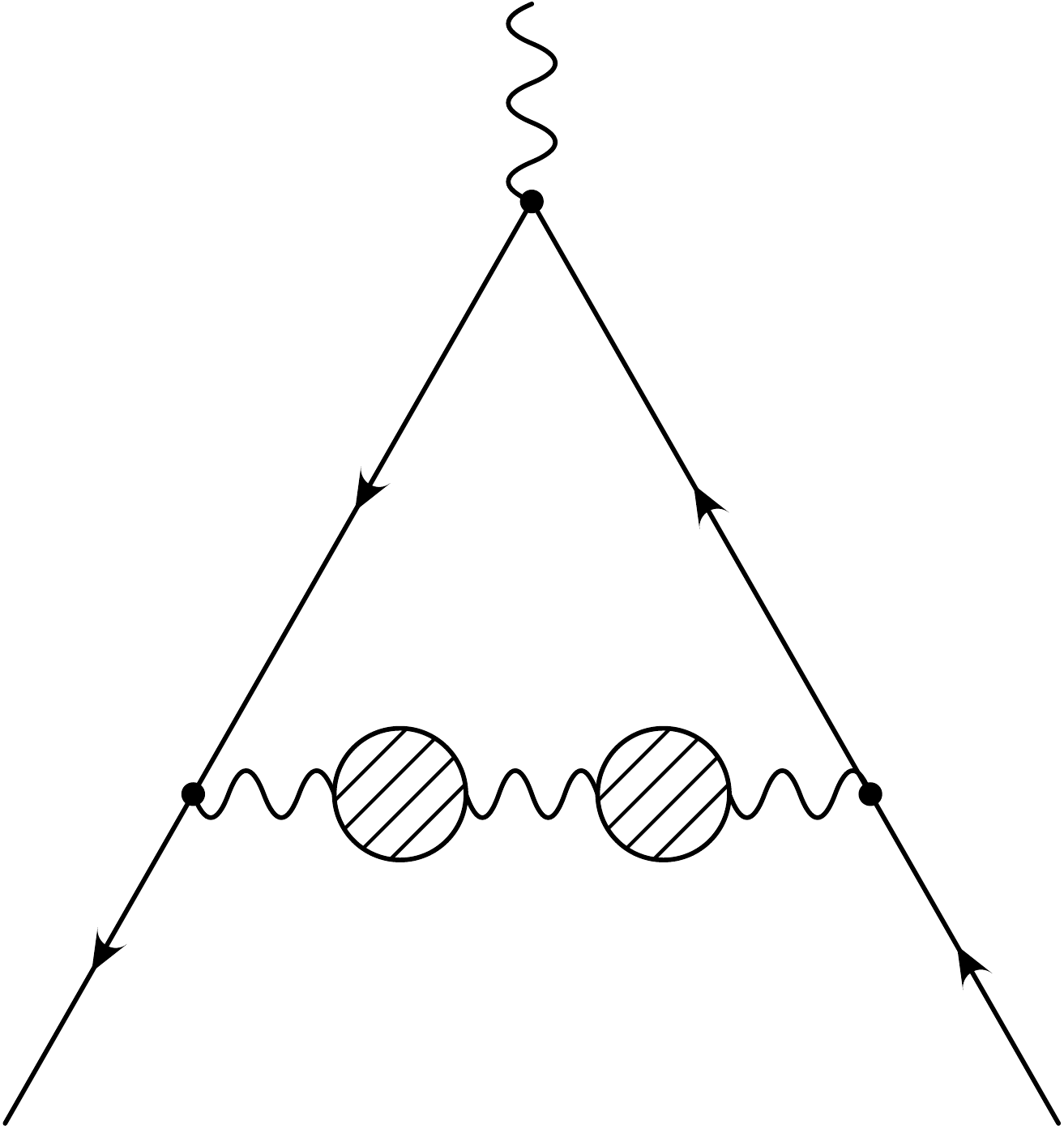}\quad
    \includegraphics[scale=0.3]{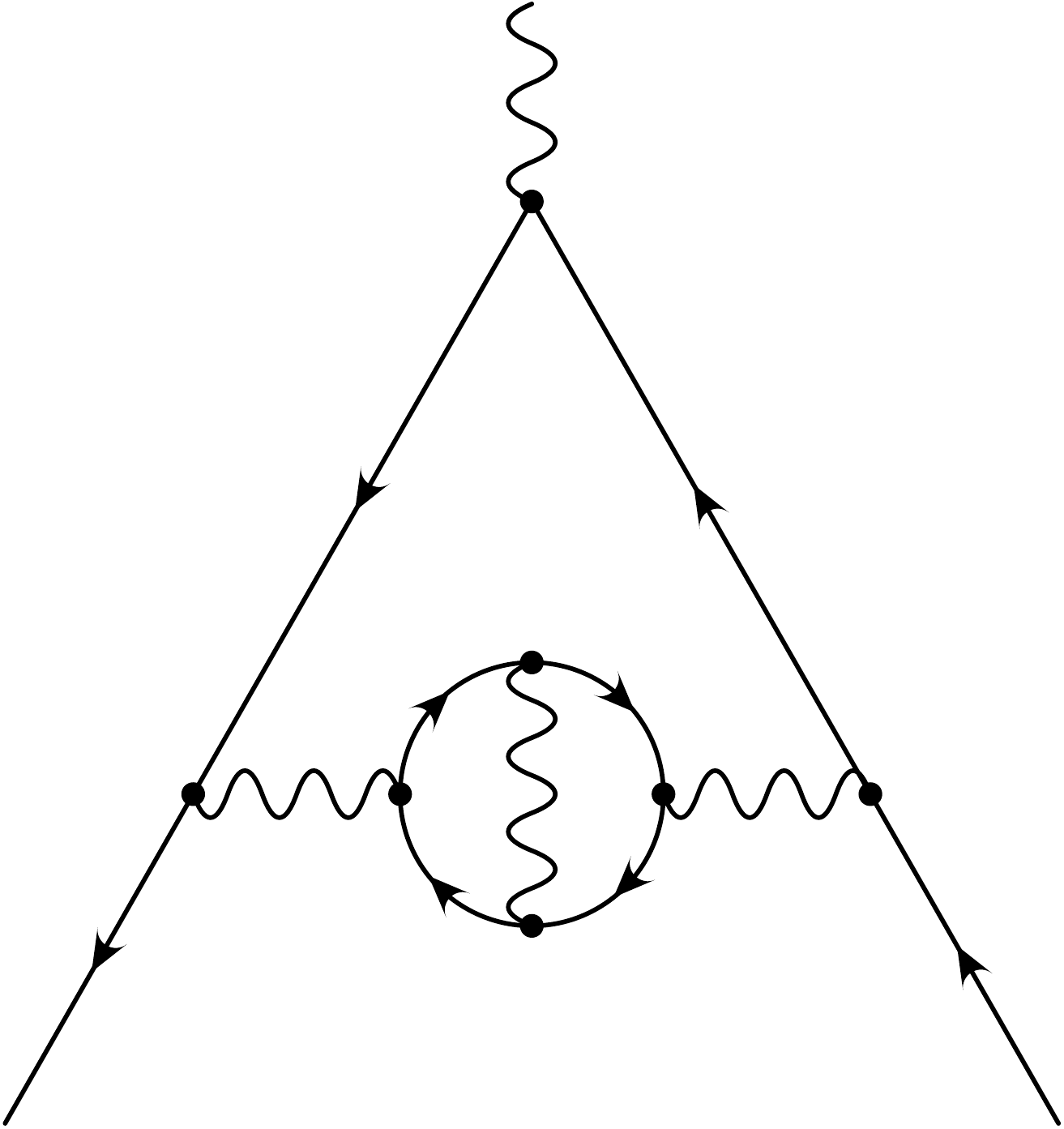}\quad
    \includegraphics[scale=0.3]{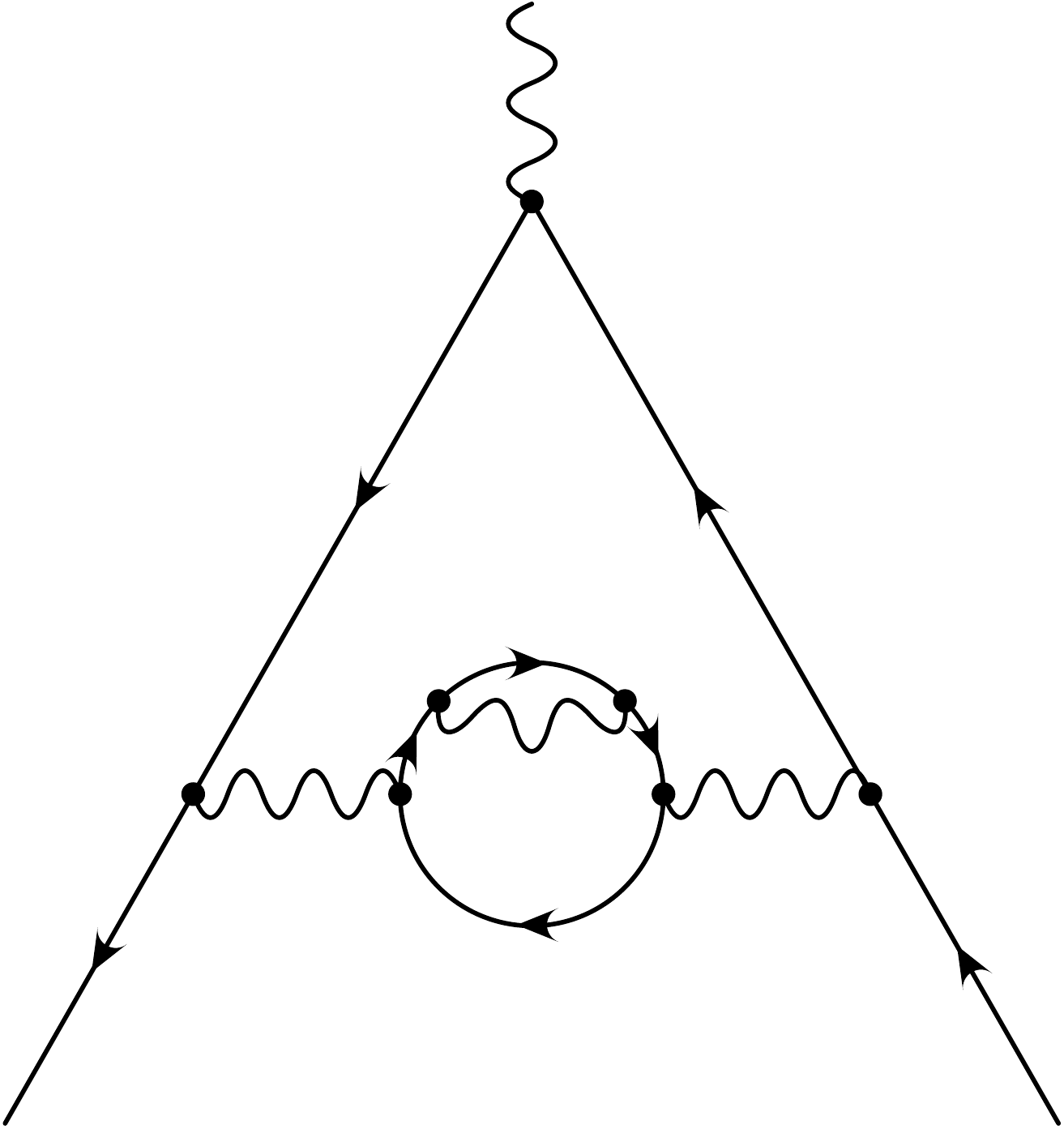}
\caption{\label{kfdiag0}Exotic examples of
  the four-point function configurations:
  double bubble and corrections to the internal structure
}
      \end{center}\end{figure}

The contributions of the four-point function $\Pi_4$ to the muon
magnetic moment is difficult to estimate
since it is completely nonperturbative by nature.
Therefore it is clearly dangerous 
to identify the corresponding
hadronic matrix element with perturbative diagrams.
We nevertheless find it convenient to characterize
different contributions at NLO by a 
decomposition of the four-point function as 
\begin{eqnarray} \label{4fact} 
&& \langle 0 | T[j(x) j(y) j(z) j(w)  ] | 0 \rangle  = 
\langle 0 | T[j(x) j(y) j(z) j(w)  ] | 0 \rangle_{\rm conn} \\ 
  \nonumber
&& \qquad \qquad + 
\langle 0 | T[j(x) j(y) ] | 0 \rangle \langle 0 | T[ j(z) j(w)  ] | 0 \rangle 
+ \mbox{permutations } x,y,z,w 
\end{eqnarray}
where $j(x)$ is a hadronic electromagnetic
current (we suppress the Lorenz index for
simplicity) and the matrix elements are to be
computed with the QCD interaction only, 
e.\,g. on the lattice. This expression is similar to the definition of
the contributions
which are one-particle irreducible with regards to the photon lines.
In perturbation theory
the second term generates double insertions of
the hadronic two-point functions
while the first term is a genuine ``nonfactorizable'' contribution to
the four-point function of electromagnetic currents. 
In the data-based analysis the two-point functions are taken from data
and 
the QED corrections that are formally given by the four-point function
are already 
partly included in the parametrization. In Fig.~(\ref{kfdiag0})
we illustrate this situation by contributions of
the leading order perturbative diagrams. The QED corrections to
the connected part which is given by the light-by-light configuration
at the leading order will generate 
terms beyond NLO in the fine structure constant
and should be dropped.  

While our modeling of hadronic contributions
may look a bit na\"ive and certainly has not yet been rigorously
justified in our analysis
as a systematic effective theory
the method is physically attractive and
very efficient computationally. Indeed, theoretically  
the four-point function contributions can be uniquely
identified in perturbation theory
and, practically, the numerical results
can be obtained by using directly
the formulas for NLO leptonic contributions to the muon magnetic
moment
that are readily available in the literature.

    We have the following contributions: 
\begin{itemize}
\item[i)]The double vacuum polarization in perturbation theory with different
quarks is mainly analogous to the mixed lepton--quark vacuum polarization as
it does not require internal corrections to the quark loop
(K\"all\'en--Sabry correction~\cite{Kallen:1955fb}, cf.\ the next item).

The charm quark together with light modes gives
\begin{equation}
\label{dbcuds}
a_\mu^{\rm had}({\rm db};{\rm NLO};c\& uds)=0.1\times 10^{-11}\, .
\end{equation}

The reiteration of light modes with different quarks reads
\begin{equation}
\label{dbuds}
a_\mu^{\rm had}({\rm db};{\rm NLO};uds\& u'd's')=3\times 10^{-11}\, .
\end{equation}

\item[ii)]In addition to double bubbles of the same fermion we have diagrams
with an internal structure, the K\"all\'en--Sabry correction. 
In our approach the new
contribution, i.e.\ an internal structure of an effective quark loop in
$\Pi_4$, is computable. The general formula for the contributions of a fermion (without
symmetry and group factors) is given by
\begin{equation}
a_\mu^{\rm ferm}({\rm 4};{\rm NLO};{\rm ferm})
=\frac{41}{486}\left(\frac{m_\mu}{m_{\rm ferm}}\right)^2\left(\alp\right)^3\, .
\end{equation}

The contribution of charm quarks is negligible,
\begin{equation}
a_\mu^{\rm had}({\rm 4};{\rm NLO};c)= 0.3\times 10^{-11}\, .
\end{equation}

\item[iii)] The result for the
  double bubble with light quarks, where the same quark is 
  running in the loop
  is taken together with the K\"all\'en--Sabry type correction to the
single bubble and yields
\begin{equation}\label{eq:q4}
a_\mu^{\rm had}({\rm 4};{\rm NLO};uds)= 25\times 10^{-11}\, .
\end{equation}
In fact, one has to add terms with the color structure $N_c^2$ not separable
in a lepton-type calculation of Ref.~\cite{barbieri}. However, these terms can
be computed explicitly. An additional group factor is
$N_c(Q_u^4+Q_d^4+Q_s^4)=2/3$, and the result is small, $0.5\times 10^{-11}$.
  
The NLO estimates of this type based on data are named ``dispersive NLO'' and are defined 
to be any contribution except the genuine light-by-light piece. 
The result is~\cite{PDG} 
\begin{equation}
a_\mu({\rm disp};{\rm NLO};{\rm had};e^+e^-)=-98.7(0.9)\times 10^{-11}\, .
\end{equation}
This numerical value
corresponds to the contribution of the two-point hadronic
function and should be compared to the sum of our results above.

\item[iv)]The light-by-light contribution is the genuine $\Pi_4$
contribution that is most unknown and controversial. Unlike for the hadronic two-point 
function there is no way at present
to extract sufficient helpful information about this contribution
from experiment and thus this topology contributes
a large part of the theoretical uncertainty. 

We use our approach described in
Sect.~\ref{sec:technique} and compute the LBL in terms 
of a ``dressed'' quark with the effective
mass $m^*$. The LBL contribution for fermions 
then reads~\cite{lblanal}
\[
a_\mu^{\rm ferm}({\rm LBL};{\rm NLO};m_q)=\left(\alp\right)^3\left\{
\left(\frac{m_\mu}{m_q}\right)^2
\left(\frac{3}{2}\zeta(3)-\frac{19}{16}\right)
\right.
\]
\begin{equation}
\left.
+
\left(\frac{m_\mu}{m_q}\right)^4\left(
-\frac{161}{810}\ln^2\left(\frac{m_q}{m_\mu}\right)
-\frac{16189}{48600}\ln\left(\frac{m_q}{m_\mu}\right)
+\frac{13}{18}\zeta(3)-\frac{161}{9720}\pi^2
-\frac{831931}{972000}\right)\right\}\, .
\end{equation}
Multiplying with the necessary group factors we can compute the light-by-light 
contribution for the different fermions.  
The light quarks give the contribution
\begin{equation}\label{eq:lbl-uds}
a_\mu^{\rm mod}({\rm LBL};{\rm NLO};uds)=139\times 10^{-11}
\end{equation}
while the result for the $c$ quark is
\begin{equation}\label{eq:lbl-c}
a_\mu^{\rm mod}({\rm LBL};{\rm NLO};c)=2\times 10^{-11}\, .
\end{equation}
We point out that this estimate is model dependent, so we also cannot assign 
a reliable estimate of the corresponding theoretical uncertainty.  
\end{itemize}

\section{Results\label{sec:res}}
We are now ready to collect the different contributions. 
The NLO result related to the two-point function is the sum of
Eqs.~(\ref{eq:NLO-ver}), (\ref{eq:e-uds}) and~(\ref{eq:e-c}),
\begin{equation}
a_\mu^{\rm had}({\rm NLO};\Pi_2)
=(-192+104+1)\times 10^{-11} =-87\times 10^{-11}\, .
\end{equation}
This result is based on LO from Eq.~(\ref{eq:LOuds-data}) and is very stable.
Our model calculation reproduces the integration of the LO data with
appropriate kernels. This is because the two kernels $K^{(2)}$ and $K^{(4)}$
behave similarly in the important region of integration (cf.\ e.g.\
Ref.~\cite{Groote:2001vu}).

The result related to the four-point function without LBL is given by
Eqs.~(\ref{dbuds}) and~(\ref{eq:q4}) and by some other small terms
($1\times 10^{-11}$),
\begin{equation}
a_\mu^{\rm had}({\rm NLO};\Pi_4{\rm pol})=(3+25+1)\times 10^{-11}
=29\times 10^{-11}\, .
\end{equation}
The main contribution comes from the K\"all\'en-Sabry term~(\ref{eq:q4}).
In fact, one would perhaps had to subtract this term from the LO contribution
before fitting $m^*$. However, as we have already discussed before,
the problem of double counting is too complicated to be considered here. 

The LBL term from Eqs.~(\ref{eq:lbl-uds}) and~(\ref{eq:lbl-c}) gives
\begin{equation}
a_\mu^{\rm had}({\rm NLO};\Pi_4{\rm LBL})
  =(139+2)\times 10^{-11}=141\times 10^{-11}\, .
\end{equation}
Therefore, the NLO contribution related to the four-point function is given by
\begin{equation}
a_\mu^{\rm had}({\rm NLO};\Pi_4{\rm tot})
  =(141 + 29)\times 10^{-11}=170\times 10^{-11}\, .
\end{equation}
A difficult question is to estimate the accuracy of the obtained
results.
It is
clear that the statistical uncertainty due to the error of the only parameter
of our model $m^*=180\pm 0.5\MeV$ is very small and is basically
irrelevant while the main uncertainty of the predictions
is a systematic one, i.e.\ the uncertainty of the model itself. As we have
discussed in detail in Sects.~\ref{sec:technique} and~\ref{sect:disc},
we think that the actual contribution of the four-point function can be up to
a factor two larger. We take a
conservative point of view and include a 50\%
uncertainty in our result as a systematics to get the prediction for the NLO
contribution related to the four-point function in the form
\begin{equation}\label{eq:four-point}
a_\mu^{\rm had}({\rm NLO};\Pi_4{\rm fin})=(170\div 255)\times 10^{-11}
=(213\pm 43)\times 10^{-11}\, .
\end{equation}
Unfortunately, the conservative uncertainty of the prediction in
Eq.~(\ref{eq:four-point}) is larger than an allowed uncertainty of
$\sigma_{\rm future}\approx (10\div 15)\times 10^{-11}$, but we believe that
based on the current techniques it is a realistic one.

Our prediction for the total hadronic NLO now reads
\begin{equation}
a_\mu^{\rm had}({\rm had};{\rm NLO})
=[-87+(213\pm 43)]\times 10^{-11}=(126\pm 43)\times 10^{-11}\, .
\end{equation}
The data-based result is 
\begin{equation}
a_\mu^{\rm had}({\rm N(N)LO};e^+e^-) = 19(26)\times 10^{-11}\, .
\end{equation}
In both cases the error is dominated by the LBL contribution, or,
more generally, by
the contribution of the four-point hadronic function.

Using the hadronic LO contribution from Eq.~(\ref{eq:pdgLO}) we obtain the
total hadronic contribution
\begin{equation}
a_\mu^{\rm had}({\rm had})
=[(6931\pm 34)+(126\pm 43)]\times 10^{-11}=(7057\pm 55)\times 10^{-11}\, .
\end{equation}
Adding in the leptonic contributions we obtain the SM value  
\begin{equation}
a_\mu^{\rm SM}=116\,591\,929(55)\times 10^{-11}\, .
\end{equation}
and the comparison with the data yields
\begin{equation}
a_\mu^{\rm exp}-a_\mu^{\rm SM}
  =\Delta a_\mu ({\rm SM})=162(54)(33)(55)\times 10^{-11}\, .
\end{equation}
showing the well known tension at the level of 
$2\sigma$'s. A future measurement with a significantly reduced uncertainty 
will certainly shed some light on this
tension, but the theoretical uncertainty
seems to persist to stay
larger than
the experimental one. The lattice calculations can become crucial for
the analysis of this observable in the SM  with such a high
accuracy.
\section{Discussion\label{sect:disc}}
We have revisited and discussed
the NLO hadronic contributions to the muon anomalous 
magnetic moment.
The contribution originating from the hadronic two-point function 
is reasonably well under control
since the bulk part of it can be extracted from the data on 
$e^+ e^-$ scattering to hadrons.
On the other hand, for the theory
an important observation is that this contribution
can be computed using the two-point function of quark electro-magnetic
currents in Euclidean domain. While this route is more reliable for
performing theoretical calculations than using data
the main obstacle persists --
the contribution of light quarks in perturbation
theory is infrared divergent.
In our model we cure the
deficiency of the perturbation theory approach 
by the direct method of regularizing the emerging infrared divergence
by introducing
an effective mass for a light quark.
It is not a formal regularization
parameter 
though but a quantity with solid physical meaning.
It is well known that generation
of a mass parameter  at small momenta
for the originally massless Lagrangian
quark is a general feature of QCD where the
chiral symmetry is spontaneously broken. As a technical implementation
of this physical feature in our method
we choose the effective mass to be a constant
as the relevant observable is an integral over the momenta that is
saturated at scales around one GeV. The numerical value for this
constant is fixed from the results of LO analysis based on data.
However, using data for computing the LO contribution
requires a careful analysis of which 
pieces are included in the data input and which need to be subtracted.
A na\"ive analysis 
based on Feynman diagrams bears the danger of double counting,
e.g. the contributions 
from the double-bubble diagrams. 

The important problem is also related to the input of the fine-structure
constant $\alpha$ at the NLO analysis. 
At the current level of precision one can use the accurately
measured value of $\alpha$ 
from the electron anomalous magnetic moment
which eventually means that we compare 
the anomalous magnetic moments of the muon and the electron.
A lot of the  
uncertainties are common to
both which also means that the uncertainties in both 
quantities are correlated even though the mass dependence of the
results is much more important in the muon case.
A similar remark may apply to other precise sources for extracting $\alpha$. 

The main theoretical
problem in the context of hadronic contributions to the muon magnetic
moment is the reliable predictions for the 
hadronic four-point 
function. In the paper we discussed a model
based on considerations of chiral symmetry 
breaking.  
As an extension of the simple version of the model of
ref.~\cite{Pivovarov:2001mw} we now
argue that the contributions related to the hadronic
four-point function can be
essentially enhanced. This conclusion emerges from the
analysis of the large number of different topologies
that appear in higher orders of perturbation theory and can play an
important role in lattice computations.
It is possible that the expectations based on perturbation theory
in Fock space with a small number of low mass resonances give an
oversimplified picture of a genuine four-point function contribution.
Indeed, in the analysis inspired by
data the main contribution to LBL
comes from an exchange by the neutral pion due to
the anomalous dimension-five interaction
$\pi^0{\tilde G}G/f_\pi$~\cite{pseudo}
(see also Refs.~\cite{Dorokhov:2012qa,Dorokhov:2015psa,Dorokhov:2011zf}).
However, within the chiral perturbation theory ($\chi$PT) approach this
contribution is subleading in power counting and, therefore, not
unique.
It is suppressed by a natural
$\chi$PT scale $\Lambda_\chi = 2\pi f_{\pi}\approx m_\rho$. The leading
contribution given by the Goldstone modes is very small. Thus, the
charged pions give~\cite{Kuhn:2003pu}
\begin{eqnarray}
&&a_\mu({\rm LBL})=\left(\alp\right)^3 a_\mu(\gamma\gamma);\nonumber \\
&&a_\mu(\gamma\gamma;{\rm SQED})=\frac{m_\mu^2}{m_\pi^2}
\left(\frac{1}{4}\zeta(3)-\frac{37}{96}\right)=-0.0849
\frac{m_\mu^2}{m_\pi^2}\, ,
\end{eqnarray}
and the contribution of charged kaons is totally negligible. Note also that
the sign is negative compared to fermions,
\begin{equation}
a_\mu(\gamma\gamma;{\rm QED})=\frac{m_\mu^2}{m_f^2}
\left(\frac{3}{2}\zeta(3)-\frac{19}{16}\right)=0.6156
\frac{m_\mu^2}{m_f^2}\, .
\end{equation}
The smallness of the pions contribution is related to the fact that pions are
spinless particles with no own magnetic moments. In view of $\chi$PT
power counting,
however, the contribution of vector mesons like $\rho$-mesons, or even of
baryons like protons, are of the same order as the neutral pion contribution,
since the scales are close ($m_\rho=777\MeV$ and 
$\Lambda_\chi=600\div 800\MeV\approx 2\pi f_\pi$). Within various effective
theory schemes, even neutrons can contribute as they interact with photons via
their own magnetic moment. Therefore, there are many contributions that are
formally of the same order in power counting
as the neutral pion one. The contribution of
$\pi^0$ in its local form is ambiguous as it depends strongly on the
ultraviolet cuts used and the usual cut is provided by the $\rho$-meson mass.

Therefore, the contributions related to the four-point function can be
enhanced even in the resonance-based approach.
It supports the conclusion obtained
by looking at the number of different
topologies that emerge at higher orders of perturbation theory. And even
though the perturbative QCD is not applicable for their quantitative evaluation,
they all appear in the analysis within the lattice approach.

To conclude, we think that there is still
some room in the SM to accommodate for the
current experimental value of the muon anomalous magnetic moment, and we are
looking forward to results of new measurements.

\subsection*{Acknowledgments}
This research was supported by the Deutsche Forschungsgemeinschaft 
(DFG, German Research Foundation) under grant  396021762 - TRR 257 
``Particle Physics Phenomenology after the Higgs Discovery''. The research of
S.G.\ was supported by the European Development Fund
under Grant No.~TK133.

\begin{appendix}

\section{Appendix\label{sec:appendix}}
\setcounter{equation}{0}\def\theequation{A\arabic{equation}}

\subsection{The LO hadronic contribution phenomenology}
The two-point correlator is given by
\be
\label{vacpolin}
i\int \langle Tj_\mu^{\rm had}(x)j_\nu^{\rm had}(0) \rangle e^{iqx}dx=
(q_\mu q_\nu -g_{\mu\nu}q^2)\Pi^{\rm had}(q^2)\, .
\ee
The definition of the fine structure constant
$\alpha$ requires a normalization $\Pi^{\rm had}(0)=0$.
The dispersion representation for  $\Pi^{\rm had}(q^2)$
can then be written with one subtraction,
leading to
\begin{equation}
\label{disprelhad}
\Pi^{\rm had}(q^2)=\frac{q^2}{\pi}\int_{4m_\pi^2}^\infty
\frac{ds}{s}\frac{{\rm Im}~\Pi^{\rm had}(s)}{s-q^2}\, .
\end{equation}
The LO expression for the muon anomaly is
\begin{equation}
\label{directcont}
a_\mu^{\rm had}({\rm LO})
=4\pi\left(\frac{\al}{\pi}\right)^2\int_{4m_\pi^2}^\infty
\frac{ds}{s}K^{(2)}(s){\rm Im}~\Pi^{\rm had}(s)
\end{equation}
with a one-loop kernel of the form
\begin{equation}
\label{eq:k2}
K^{(2)}(s)=\int_0^1dx\frac{x^2(1-x)}{x^2+(1-x)s/m_\mu^2}.
\end{equation}
This expression is useful for the analysis based on the hadronic cross section
of $e^+e^-$ annihilation. 

For the theory analysis, Eq.~(\ref{directcont}) can be rewritten as an
integral over Euclidean values $t=-q^2$ for $\Pi^{\rm had}(q^2)$,
\begin{equation}
\label{directcontEucl}
a_\mu^{\rm had}({\rm LO})=4\pi^2\left(\frac{\al}{\pi}\right)^2
\int_0^\infty \left\{-\Pi^{\rm had}(-t)\right\} W(t)dt
\end{equation}
with
\begin{equation}
W(t)=\frac{4m^4}{\sqrt{t^2+4m^2t}\left(t+2m^2+\sqrt{t^2+4m^2t}\right)^2}.
\end{equation}
This form is well known in the form of a parametric
integral~\cite{samuel,leadorder}.

Eq.~(\ref{directcontEucl}) is more suitable for a theoretical study as the
theory is preferably applicable in Euclidean domain and on the lattice in
particular. One can further write
\begin{equation}
\label{ffuncparts}
\frac{1}{\pi}\int_{4m_\pi^2}^\infty
\frac{ds}{s}K^{(2)}(s){\rm Im}~\Pi^{\rm had}(s)=
\int_0^\infty \left(-\frac{d \Pi^{\rm had}(-t)}{d t}\right) F(t)dt,
\qquad
F(t)= \int_t^\infty W(\zeta)d\zeta
\end{equation}
with
\begin{equation}
\label{expliffun}
F(t)=\frac{1}{2}\left(
\frac{t+2m^2-\sqrt{t^2+4m^2t}}{t+2m^2+\sqrt{t^2+4m^2t}}\right)
=\frac{2m^4}{\left(t+2m^2+\sqrt{t^2+4m^2t}\right)^2}.
\end{equation}

Therefore, the analysis of the anomaly is based on the derivative of the
hadron vacuum polarization function $d\Pi^{\rm had}(-t)/dt$ which is closely
related to the famous Adler function~\cite{Adler:1974gd}
\begin{equation}
D(t)= -t\frac{d \Pi^{\rm had}(-t)}{dt}.
\end{equation}
The Adler function can be computed in perturbative QCD with massless
quarks for large $t$,
\begin{equation}
D(t) = e_q^2 N_c \frac{1}{12\pi^2}
  \left(1+\frac{\al_s(t)}{\pi}\right)\, .
\end{equation}

\subsection{The LO: technical formulas}
The LO technical formulas given here are used to
fix $m_q$ from the LO hadronic contribution.
A fermion with mass $m_q$ without QCD group factors (as a lepton) gives a LO
contribution to the muon anomaly of the form
\begin{equation}
\label{mod0vac}
a_\mu^{\rm ferm}({\rm LO})
= I(m_q)\left(\frac{\al}{\pi}\right)^2
\end{equation}
with
\begin{equation}
I(m_q)=\int_{4m_q^2}^\infty \frac{\rho_q(s)K^{(2)}(s)}{s}ds
\end{equation}
and
\begin{equation}
\label{fermspect2}
\rho_q(s)=\frac{1}{3}
\sqrt{1-\frac{4 m_q^2}{s}}\left(1+\frac{2 m_q^2}{s}\right)\, .
\end{equation}
The explicit integration over $s$ with the kernel $K^{(2)}(s)$ from 
Eq.~(\ref{eq:k2}) gives
\begin{equation}
\label{intrepI}
I(m_q)=\int_0^1 dx (1-x) [-\pi(x,m_q)]
\end{equation}
where 
\begin{equation}
\pi(x,m_q)=\left(\frac{1}{3z}-1\right)\varphi(z)-\frac{1}{9}
\end{equation}
and
\begin{equation}
\label{lasttech2}
\varphi(z)=\frac{1}{\sqrt{z}}{\rm artanh}(\sqrt{z})-1,\quad
z=\frac{m_\mu^2 x^2}{4m_q^2(1-x)+ m_\mu^2 x^2}
  =\frac{t}{4m_q^2+ t}\, ,\quad t=\frac{m_\mu^2 x^2}{(1-x)}\, .
\end{equation}
An analytical expression for the function $I(m_q)$ is known. However, the
integral representation given in Eq.~(\ref{intrepI}) is sufficient for
practical applications.

The iterated contribution for two fermions (double bubble generalization of
Eq.~(\ref{intrepI})) is given by~\cite{samuel}
\begin{equation}
a_\mu^{\rm ferm}({\rm db};f_1\& f_2)=\left(\alp\right)^3
\int_0^1 dx (1-x) (-\pi(x,m_{f_1}))(- \pi(x,m_{f_2}))\, .
\end{equation}
The actual application of this formula in QCD should account for symmetry
factors (a factor $2$ if the fermions are different) and for group factors.

\end{appendix}

\end{document}